\pdfoutput=1

\documentclass[entropy,article,accept,moreauthors,pdftex,12pt,a4paper]{mdpi}
\usepackage{graphicx}
\usepackage{amsmath}
\usepackage{bbm}
\usepackage{amsfonts}
\usepackage{amssymb}

\lastpage{1210}
\doinum{10.3390/e16031191}
\pubvolume{16}
\pubyear{2014}
\history{Received: 10/Sep/2013 / Published: 25/Feb/2014}

\Title{One antimatter --- two possible thermodynamics}

\Author{A. Y. Klimenko$^{1,*}$ and  U. Maas $^{2}$}

\address{%
$^{1}$ The University of Queensland, SoMME, QLD 4072, Australia\\
$^{2}$ Karlsruher Institut fur Technologie, ITT, 76131, Germany}

\corres{E-mail: klimenko@mech.uq.edu.au}

\abstract{Conventional thermodynamics, which is formulated for our world populated by radiation and matter, can be extended to describe physical properties of antimatter in two mutually exclusive ways: CP-invariant or CPT-invariant. Here we refer to invariance of physical laws under charge (C), parity (P) and time reversal (T) transformations. While in quantum field theory CPT invariance is a theorem confirmed by experiments, 
the symmetry principles applied to macroscopic phenomena or to the whole of the Universe represent only hypotheses.  
Since both versions of thermodynamics are different only in their treatment of antimatter, but are the same in describing our world dominated by matter, making a clear experimentally justified choice between CP invariance and CPT invariance in context of 
thermodynamics is not possible at present. This work investigates the comparative properties of the CP- and CPT-invariant extensions of thermodynamics (focusing on the latter, which is less conventional than the former) and examines conditions under which these extensions can be experimentally tested.  
}


\begin{document}

%
%
%
%
%
%
%
%
%
%
%
%
%
%
%
%
%
%
%
%
%
%
%
%
%
%
%
%
%
%
%
%
%
%
%
%
%
%
%
%
%
%
%
%
%
%
%

\section{Introduction}

In the 1890s, the kinetic theory of Ludwig Boltzmann, which represents an
important link between thermodynamics and classical mechanics, attracted both
interest and criticism. The criticism was to some extent motivated by doubts
about the atomic (molecular) structure of matter, which were quite persistent
at that time, but also involved a series of very interesting questions about
consistency of the reversibility of classical mechanics with the irreversible
nature of thermodynamics. Some of these questions (e.g. the exact physical
mechanism determining the direction of time) are not fully answered even
today. In response to his critics, Boltzmann put forward a number of
hypotheses of remarkable originality and depth
\cite{Boltzmann-nature,Boltzmann-book}. One of these hypotheses identifies our
perceived direction of time with the second law of thermodynamics. Another
hypothesis links the second law to a giant fluctuation and the temporal
boundary conditions imposed on the Universe (or the observed part of it). The
consequence of these hypotheses is the astonishing possibility (which was
explicitly discussed by Boltzmann \cite{Boltzmann-book}) that, given different
temporal boundary conditions, the perceived time may run in opposite
directions in different parts of the Universe.

More than 100 years after, we still do not have a full explanation for the
physical mechanism of the direction of time, and the second law of
thermodynamics remains our key indicator for the time arrow. Our modern
understanding of the Universe is, however, quite different from the views
common in the late 19th century \cite{PenroseBook}. While giant thermodynamic
fluctuations or radically different initial conditions imposed on different
parts of the Universe seem rather unlikely, Boltzmann's thermodynamic time
running in different directions might still be possible due to (potential)
existence of antimatter. Since Feynman's theory of positron \cite{Feynman1949}%
, antiparticles are seen as particles moving backward in time. This gives us a
hint that conventional thermodynamics, which is formulated for our world full
of radiation and matter but, as far as we know, free of assembled antimatter,
can be extended to antimatter in two possible ways CP-invariant or
CPT-invariant. Our discussion is aimed at the macroscopic properties, while
the effect of thermodynamic interference on invariant properties of quantum
systems is discussed elsewhere \cite{K-PhysA}.

While Sakharov \cite{Sakharov1967} suggests that CPT symmetry can be a global
property of the Universe, Penrose \cite{PenroseBook} believes that CPT
symmetry does not hold on such scales due to\ the action of the second law.
The competing physical intuitions of CPT invariance and of conventional
thermodynamics have been recently discussed by Downes et al \cite{CPT2012}.
The present approach should not be confused with the analysis of isoduality of
thermodynamic equations conducted by Dunning-Davies \cite{Dunning-Davies} on
the basis of Santilli isodualities and having physical outcomes and
interpretations very different from ours. Our work treats matter and
antimatter as being the same with respect to the 1st law of thermodynamics
(i.e. all reversible and mechanical laws), does not change the signs of energy
and mass, and does not imply existence of antigravity and anti-photons that
are associated with Santilli isodualities. A popular presentation of
CPT-invariant thermodynamics is given in Ref.\cite{K-M-2012} from the
perspective of a space traveller visiting the world made of antimatter.

Quantum decoherence is commonly seen as the process initiating entropy
increases, which are then reflected in macroscopic irreversibilities of
temporal evolutions. Two approaches to decoherence can be distinguished:
induced and spontaneous (intrinsic). The induced decoherence occurs within
unitary description of quantum systems but involves interference of the
environment \cite{Zurek1982,CT-P2009,Yukalov2012}. The spontaneous decoherence
implies violations of unitarity that can be very small in magnitude to be
detected directly but the overall effect these violations is amplified in
quantum or mechanical systems of large dimensions and appears to be profound
--- the macroscopic increase in entropy \cite{Zurek2002,
QTreview,Beretta2005,PenroseBook}. Various possible causes and mechanisms of
decoherence can be nominated but, since the exact microscopic physical
mechanisms enacting thermodynamic time are still not known, we use the term
\textit{time primer} as the place holder for these mechanisms without
referring to any specific theory. It seems that both mechanisms, induced and
intrinsic, are likely to coexists in the real world and can be referred to
here as induced and intrinsic priming of time. Since our consideration
pertains to the thermodynamic perspective, the details related to priming the
direction of time are discussed in the Appendix. The two versions of
thermodynamics considered here correspond to the intrinsic time priming having
the same (CP-invariant thermodynamics) or opposite (CPT-invariant
thermodynamics) temporal directions for matter and antimatter.

Microscopic symmetries have been repeatedly discussed in publications in the
context of particle physics \cite{Symmetry1976,PDG2012}. The current status of
the experimental confirmation of quantum symmetry principles is that C, P, T
and CP symmetries are broken by weak interactions but the CPT symmetry, which
is linked to the fundamental Lorentz invariance, is believed to be upheld
\cite{PDG2012}. The CP and T violations but are much less common than
CP-preserving C and P violations. For a many decades the CP violation observed
in decays of neutral K-mesons was deemed to be the only known exception
\cite{CP1964,Symmetry1976}. The discoveries of CP violations for B-mesons by
BABAR\cite{BABAR2001}, Belle \cite{Belle2001} and LHCb \cite{LHCb2013Bs0}
collaborations indicate that CP violations are rare but not exceptional,
although there is no reliable case of a\ CPT violation known in particle
physics \cite{PDG2012}. While CP violation combined with CPT invariance
implies T violation, a direct confirmation of T violation has been obtained
only very recently \cite{BABAR2012t}.

\section{Symmetry of matter and antimatter}

The Universe is populated mostly by radiation, has significant quantities of
matter and, as far as we know, very small quantities of antimatter in the form
of scattered elementary particles. Antimatter is not identical to matter but
there is a strong similarity between them \cite{Symmetry1976}. We give this
similarity a very broad interpretation: if our Universe were mostly composed
from antimatter, we (also made of antimatter) would not be able to tell the
difference as the physical laws of the antimatter universe would be the same.

If we change matter to corresponding antimatter by so called charge
conjugation or (C-conjugation), the heat fluxes $\mathbf{q}$ can change
according to the following linear operation $\mathbb{C}(\mathbf{q}%
)=\alpha\mathbf{q}$\textbf{$\mathbf{,}$} where $\mathbf{q}$ is a conventional
vector and $\alpha$ is a real constant. This operation takes us to
antiuniverse with the same physical laws as ours but the constant $\alpha$ is
not necessarily $1$ since we do not know the correspondence of space and time
between the two universes, composed of matter and antimatter. Applying the
charge conjugation again results in converting antimatter back to matter, i.e.
to going back to the original state. This means that%
\begin{equation}
\mathbf{q=}\mathbb{C}\left(  \mathbb{C}(\mathbf{q})\right)  =\mathbb{C}%
(\alpha\mathbf{q})=\alpha^{2}\mathbf{q\;\;\Longrightarrow\;}\;\alpha
=\pm1\label{CP_CPT}%
\end{equation}
These two values of $\alpha$ correspond to different thermodynamics:
C-invariant ($\alpha=+1$) and CT-invariant ($\alpha=-1$). Here we refer to
time reversal, which represents change of directions of time ($\mathbb{T}%
(t)=-t$) and thus reverses the direction of $\mathbf{q}$, as T. Considering
the parity transformation, which represents change of directions of spatial
coordinates ($\mathbb{P}(\mathbf{x})=-\mathbf{x}$) and referred to as P, we
follow accepted conventions and study CP-invariant and CPT-invariant versions
of thermodynamics (although parity is generally less interesting for us).
These operations are commonly used in quantum mechanics but here we follow
Sakharov \cite{Sakharov1967} and interpret them as general physical principles
not confined to the field of quantum mechanics. (Note that the quantum
mechanical consistency requires only that $\left|  \alpha^{2}\right|  =1,$
i.e. $\alpha^{2}$ can be complex; the classical world deals only with real
quantities and $\alpha^{2}=1$.) Sakharov's principles that are necessary for
explanation of the matter/antimatter bias are often quoted as:

\begin{itemize}
\item  existence of reactions violating baryon numbers

\item  violation of the C and CP symmetry

\item  deviation from thermodynamic equilibrium
\end{itemize}

\noindent This list, however, does not include the fourth important assumption
made by Sakharov in the same work \cite{Sakharov1967}: CPT invariance is a
global property of the Universe. While CPT invariance is commonly known as a
theorem in quantum field theory (which, as mentioned in the Introduction, is
supported by experiments conducted at microscopic level), its application to
macroscopic processes and/or to the whole of the Universe is a hypothesis, not
a theorem. For example, Penrose \cite{PenroseBook} believes that CPT
invariance would not hold if applied to the whole of the Universe\ --- this
view corresponds to the CP-invariant version of thermodynamics. The Universe,
nevertheless, can be expected to be CPT invariant if the CPT-invariant version
of thermodynamics is adopted (or if there are no irreversible processes take
place in the Universe but this does not seem to be the case).

\begin{figure}[h]
\begin{center}
\includegraphics[width=12cm,page=4,trim=4.5cm 3cm 5.5cm 0.8cm, clip ]{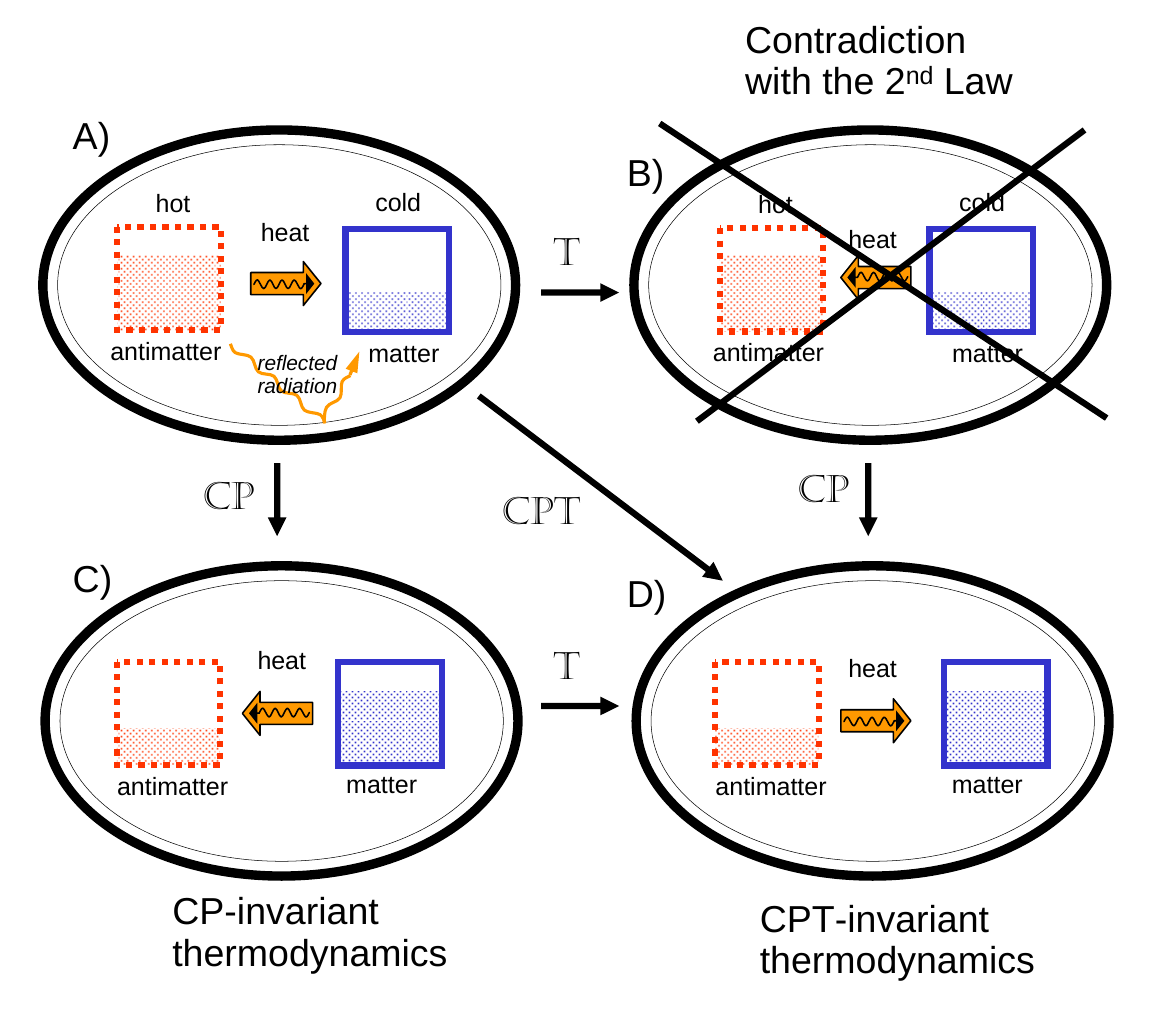}
\caption{CP and T transformations of thermodynamics systems. The fill levels  
correspond to the intrinsic temperatures. Common-sense interpretation "cold" and "hot" are indicated when applicable.}
\label{fig4}
\end{center}
\end{figure}

Since thermodynamics mainly deals with macroscopic scalar quantities, parity
transformations are, generally, not important for a thermodynamic
analysis\footnote{Although particles with opposite chiralities may need to be
treated as different species in thermodynamics and kinetics.}. Here, we
nevertheless preserve conventional notations adopted in quantum mechanics and
retain P in transformations. The implications of having positive and negative
$\alpha$ in (\ref{CP_CPT}) are shown in Figure \ref{fig4}, which illustrates
an experiment conducted by placing matter and antimatter in the focuses of a
perfectly reflective ellipsoid. One of many possible initial states, which is
shown Figure \ref{fig4}(A), is subject to the C, P and T transformations. Heat
is transferred by radiation between matter and antimatter and the direction of
the heat transfer changes as the transformations are applied. In simple terms,
T reverses the direction of the heat transfer, C converts matter into
antimatter and vice versa, while P swaps the systems located in the focuses
and reverses the heat flux accordingly. Since thermodynamic laws are not
time-invariant, the T-transformation of case (A) results in case (B)
prohibited by the second law. The other cases, (C) and (D), are possible and
correspond to CP-invariant and CPT-invariant thermodynamics. Note that, since
thermodynamics is not T-symmetric, it must also violate at least one of the
symmetries, CP or CPT. The requirements of CP-invariant and CPT-invariant
thermodynamics are thus incompatible and only one of these thermodynamics
corresponds to the real world. It is difficult, however, to determine how
thermodynamics should be extended from matter to antimatter due to the paucity
of antimatter in the Universe. Both CP- and CPT- invariant thermodynamics give
the same description for our current world populated by matter. While
CP-invariant thermodynamics is conventional and does not need an extensive
treatment here, it seems that CPT-invariant thermodynamics has not been
considered in the past and its implications need a detailed analysis.

\section{CPT-invariant thermodynamics}

Our previous discussion results in postulating the following key principles:

\begin{itemize}
\item \textbf{Reversible equivalence.} There is no distinction between matter
and antimatter with respect to the first law of thermodynamics.

\item \textbf{Inverted irreversibility.} Thermodynamically\ isolated
antimatter can increase its entropy only backward in time (unlike any isolated
matter, whose entropy increases forward in time).

\item \textbf{Observational symmetry.} Antimatter and its interactions with
matter are seen (i.e. observed, experimented with or measured) by
antiobservers in exactly the same way as matter and its interactions with
antimatter are seen by observers.
\end{itemize}

\noindent These principles correspond to CPT-invariant thermodynamics, where
the directions of thermodynamic time are opposite for matter and antimatter.
In the case of CP-invariant thermodynamics (not considered in this section),
the second principle is to be replaced by entropy increase forward in time for
both matter and antimatter. Hypothetical observers made of antimatter, are
called \textit{antiobservers,} while the term \textit{observers} refers only
to us --- observers made of matter. Properties of matter measured by us (the
observers) and properties of antimatter measured by antiobservers are referred
to as \textit{intrinsic. }The properties of matter and antimatter measured by
observers are referred to as \textit{apparent, }while the properties of matter
and antimatter measured by antiobservers are referred to as
\textit{antiapparent. }

\subsection{Interactions of thermodynamic systems and antisystems}

In this subsection, we consider thermodynamic interactions of two systems
comprised of matter and antimatter (i.e. a system and an antisystem). The
interactions are limited by the maximal energy that is allowed to be
transferred between the systems, which otherwise remain autonomous and
isolated from the rest of the Universe. Due to autonomy, we can still apply
causality in its modified form: the initial conditions for the matter system
are set before and for the antimatter system are set after the interaction.
Here, ''before'' and ''after'' refer to the observer's time.

Thermodynamics is based on determining the direction of processes where states
(i.e. macrostates) can be realised by the largest possible number of
microstates (given the constraints imposed on the system) and thus are
overwhelmingly more likely than states encompassing fewer microstates. The
logic of thermodynamics considers what is likely and neglects what is
unlikely. The most likely state is called equilibrium. In conventional
thermodynamics, unlikely states may be set as initial states while the system
tends to move towards its equilibrium as time passes. This is reflected by the
well-known Boltzmann--Planck entropy equation
\begin{equation}
S_{i}=k_{B}\ln(\Gamma_{i}) \label{BP}%
\end{equation}
linking the entropy $S_{i}$ in the state $i$ to the number of microstates
$\Gamma_{i}$ in this state. The number of microstates $\Gamma_{i}$ is further
referred to as the statistical weight of the thermodynamic state $i$. The
constant $k_{B}$ is the Boltzmann constant that rescales very large changes in
$\Gamma_{i}$ to more manageable thermodynamic quantities. In practice,
counting statistical weights of macrostates can be complicated by stochastic
dependencies. For example, if the initial state is $i_{0}$ at $t=0$ then the
most likely state $i_{\Delta t}$ at $t=\Delta t$ conditioned on the initial
state of the system may be different from the equilibrium state $i_{e}$. We
believe, however, that $i_{\Delta t}$ always becomes $i_{e}$ when $\Delta t$
is sufficiently large; i.e. stochastic dependencies disappear for states
separated by a substantial amount of time (and, possibly, space).

In this work, we do not discriminate the past and the future a priori \ ---
thermodynamic principles are applied by maximising the number of microstates
associated with macroscopic evolutions, given spatial and temporal boundary
conditions as well as other physical constraints imposed on the overall
system. The term ''overall'' stresses inclusion of both the system and the
antisystem and the temporal boundary conditions may be applied in both the
past and the future.

\subsubsection{Apparent temperatures.}

\begin{figure}[h]
\begin{center}
\includegraphics[width=12cm,page=5,trim=2.5cm 2.5cm 2.5cm 2cm, clip ]{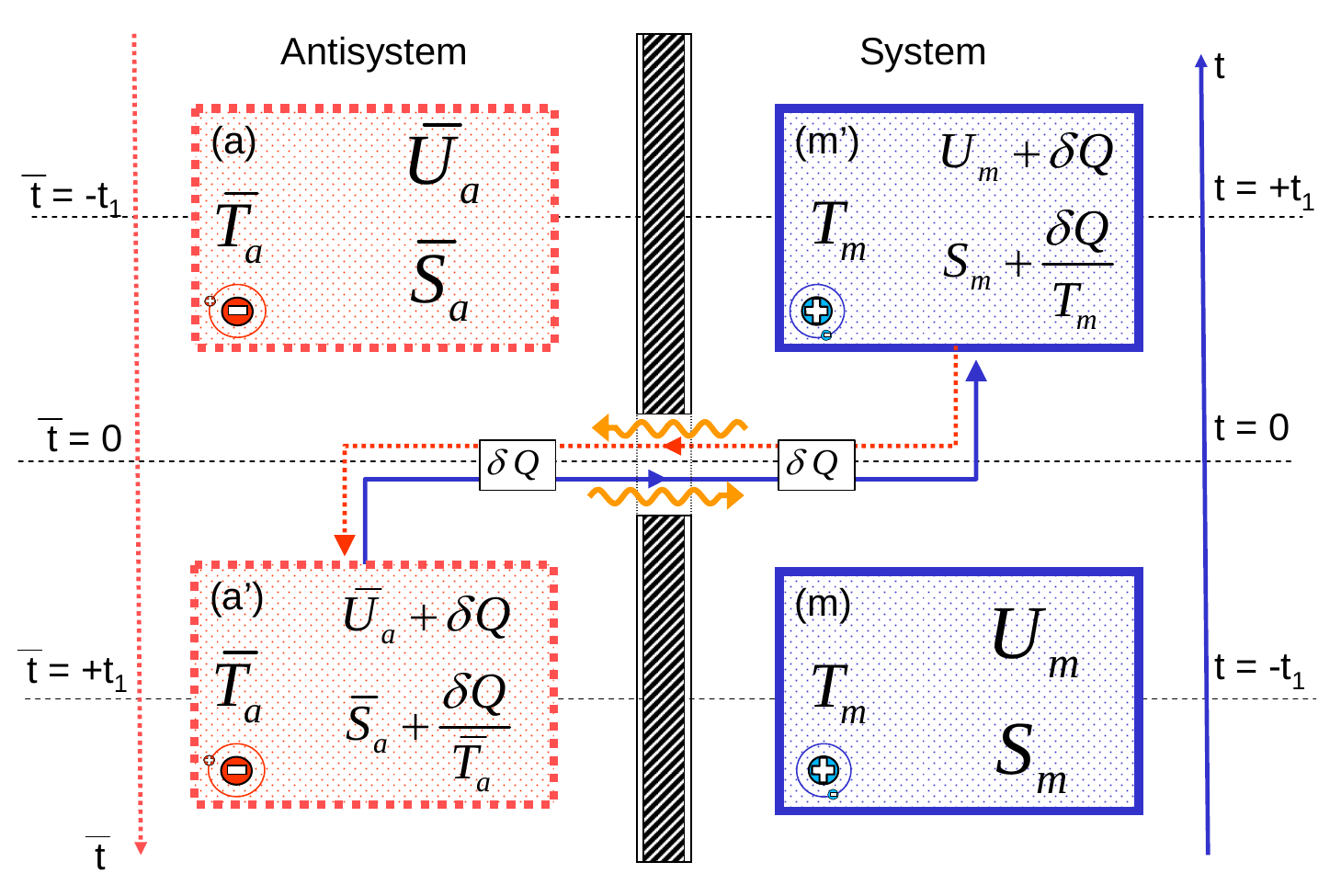}
\caption{Thermodynamic interactions of a system (right) and antisystem (left) where a limited amount of thermal energy $\delta Q < \delta Q_{\max}$ is allowed through the time window.}
\label{fig5}
\end{center}
\end{figure}

The temporal boundary conditions for the example shown in Figure \ref{fig5}
are: energy $U_{m}$ and entropy $S_{m}$ are specified for the system at
$t=-t_{1}$ and $\bar{t}=t_{1},$ that is in our past. According to inverted
irreversibility, $\bar{U}_{a}$ and $\bar{S}_{a}$ are specified for the
antisystem, at $\bar{t}=-t_{1}$ and $t=t_{1}$ that is the antisystem's
thermodynamic past and our future. The overbar symbol indicates that the value
is antiapparent, i.e. evaluated from the perspective of an antiobserver, whose
time $\bar{t}=-t$ goes in the opposite direction as compared to our time $t$.
The system and antisystem are isolated from each other for most of the time
but a limited thermodynamic contact of matter and antimatter, allowing for
transition of a small quantity of heat $\delta Q$ through exchange of
radiation, occurs at $t=0$ (and $\bar{t}=0$). The time window is selected so
that $\left|  \delta Q\right|  $ cannot exceed $\delta Q_{\max}$ where $\delta
Q_{\max}$ is sufficiently small. According to the observer the thermal energy
$\delta Q$ is transferred from the antisystem to the system as shown by the
black solid arrow. According to the antiobserver, who interprets the same
event in the opposite direction of time, the same thermal energy $\delta Q$ is
transferred from the system to antisystem as shown by the red dashed arrow.
Heat $\delta Q$ is assumed to be positive when transferred in the direction
shown in Figure \ref{fig5}: from the antisystem to the system according to the
observer and from the system to the antisystem according to the antiobserver.
The total energy
\begin{equation}
U_{tot}=U_{m}+\left(  \bar{U}_{a}+\delta Q\right)  =\left(  U_{m}+\delta
Q\right)  +\bar{U}_{a} \label{Utot}%
\end{equation}
(evaluated at any constant time $t=-\bar{t}$) is preserved in this example, as
it should since the formulation of the first law of thermodynamics does not
depend on the differences between matter and antimatter due to the postulated
reversible equivalence.\footnote{Interactions of matter and antimatter may
include the third key component --- coherent radiation. In thermodynamics,
this would correspond to considering work reservoirs in addtion to heat
reservoirs (see, for example, the ''weight process'' of Ref.\cite{Beretta1991}%
). The energy balalnce at the moment of contact is $U_{tot}=U_{m0}%
+U_{a0}+U_{r0}$ where $U_{r}$ is the radiation energy and the substript
''$0$'' \ indicates states taken at $t=0$.} \ Note that $\bar{U}_{a}=U_{a}$
due to reversible equivalence. The entropy change of the system as observed by
us and the entropy change of the antisystem as seen by the antiobserver (these
are the entropies linked to $\Gamma$) can be easily evaluated and these
changes of intrinsic entropy are shown in Figure \ref{fig5} for the states
m$^{\prime}$ and a$^{\prime}$.

We now evaluate the overall statistical weight $\Gamma_{tot}$ that corresponds
to different trajectories that are allowed by the first law of thermodynamics.
The overall state is related to the four sub-states: m, a, m$^{\prime}$ and
a$^{\prime}.$ The overall statistical weight $\Gamma_{tot}$ is linked to the
product of the statistical weights of the sub-states $\Gamma_{m}%
\Gamma_{m^{\prime}}\Gamma_{a}\Gamma_{a^{\prime}}$ and, according to equation
(\ref{BP}), becomes
\begin{equation}
\Gamma_{tot}(\delta Q)\sim\Gamma_{m}\Gamma_{m^{\prime}}\Gamma_{a}%
\Gamma_{a^{\prime}}=\exp\left(  \frac{S_{m}+S_{m^{\prime}}+\bar{S}_{a}+\bar
{S}_{a^{\prime}}}{k_{B}}\right)  \label{W1}%
\end{equation}
The value $\Gamma_{tot}$ depends on $\delta Q$. Note that $S_{m}$ and $\bar
{S}_{a}$ are fixed by the boundary conditions and only $S_{m^{\prime}}$ and
$\bar{S}_{a^{\prime}}$ depend on $\delta Q$. In this example, we should place
the time moments $t=-t_{1}$ and $t=+t_{1}$ as far apart as needed to ensure
establishment of equilibriums within the system and the antisystem before and
after the interaction, which is used in (\ref{W1}) in form of stochastic
independence of microstates that correspond to the macrostates m and
m$^{\prime}$, a and a$^{\prime}$. Two of the states, m and a, are fixed by the
boundary conditions.

Equation (\ref{W1}) can be simplified through normalising $\Gamma_{tot}$ by
the value of $\Gamma_{tot}$ at $\delta Q=0$%
\begin{equation}
\frac{\Gamma_{tot}(\delta Q)}{\Gamma_{tot}(0)}=\exp\left(  \frac{\delta
Q}{k_{B}}\left(  \frac{1}{T_{m}}+\frac{1}{\bar{T}_{a}}\right)  \right)
\label{W2}%
\end{equation}
where conventional definitions of the temperature
\begin{equation}
\frac{1}{T_{m}}=\frac{\partial S_{m}}{\partial U_{m}},\;\;\;\frac{1}{\bar
{T}_{a}}=\frac{\partial\bar{S}_{a}}{\partial\bar{U}_{a}} \label{TT}%
\end{equation}
are used. (The quantity $\delta Q$ is assumed to be too small to affect the
intrinsic temperatures of matter and antimatter, which remain $T_{m}$ and
$\bar{T}_{a}$ correspondingly.). \ From observer's perspective,\ the
conventional equilibrium condition is given by equivalence of apparent
temperatures $T_{m}=T_{a}$. If the system and the antisystem are in
thermodynamic equilibrium, both directions of heat transfer $\delta Q>0$ and
$\delta Q<0$ must be equally likely and have the same statistical weight
$\Gamma_{tot}$. This occurs only when $T_{m}=-\bar{T}_{a}$ indicating that the
apparent temperature of antimatter is $T_{a}=-\bar{T}_{a}$. In the same way,
the antiapparent (i.e. perceived by the antiobserver) temperature of matter is
$\bar{T}_{m}=-T_{m}$. It is easy to see that thermodynamic quantities $\bar
{S}_{a},\;\bar{T}_{a},\;$and $\bar{U}_{a}$ that characterise the intrinsic
properties of antimatter are apparent as%
\begin{equation}
T_{a}=-\bar{T}_{a},\;\;S_{a}=-\bar{S}_{a},\;\;U_{a}=\bar{U}_{a} \label{TSU}%
\end{equation}
from our perspective. The sign of $U_{a}$ is selected to be consistent with
the first law of thermodynamics (\ref{Utot}), while the sign of $S_{a}$ is
chosen to be consistent with the definition of temperature $T_{a}%
^{-1}=\partial S_{a}/\partial U_{a}$\ and with equations (\ref{TT}). The
change of sign does not affect our interpretation of reversible
transformations of antimatter since $S_{a}$ is constant whenever $\bar{S}_{a}$
is constant, which is consistent with our assumption that matter and
antimatter behave in the same way in reversible processes. The state of having
the same positive (and finite) intrinsic temperatures $T_{m}=\bar{T}_{a}$ does
not correspond to equilibrium and, according to (\ref{W2}), transfer of heat
from antisystem to system in observer's time is strongly favoured by thermodynamics.

It appears that a system created in our world with negative temperatures can,
at least in principle, be placed into thermal equilibrium with an antisystem
held at positive intrinsic temperatures or, analogously, an antisystem at
negative intrinsic temperatures can be in thermodynamic equilibrium with a
system having a positive intrinsic temperature. This equilibrium state,
however, would be predominantly unstable, since in most cases the antimatter
system according to (\ref{TSU}) is likely to have negative apparent heat
capacities (see, for example, Ref. \cite{K-OTJ2012})%
\begin{equation}
C_{a}=\frac{-1}{T_{a}^{2}}\left(  \frac{\partial^{2}S_{a}}{\partial U_{a}^{2}%
}\right)  ^{-1}=\frac{1}{\bar{T}_{a}^{2}}\left(  \frac{\partial^{2}\bar{S}%
_{a}}{\partial\bar{U}_{a}^{2}}\right)  ^{-1}=-\bar{C}_{a} \label{CC}%
\end{equation}
This equation indicates that changing the sign of the entropy $S$ changes the
sign of the heat capacity $C$ irrespective of the sign of the temperature $T$.
\ Hence, a thermodynamic contact of an antisystem and a system predominantly
results not in reaching the corresponding thermal equilibrium state but in the
antisystem losing energy (to the system) and further increasing its apparent
temperature until the temperature of the antisystem reaches its intrinsic
ground state $-T_{a}=\bar{T}_{a}\rightarrow+0$. Note that (neutrally) stable
thermodynamic equilibrium between a system and an antisystem is possible at
$T_{m}=T_{a}=\bar{T}_{a}=\infty$.

\subsubsection{Mass exchange between matter and antimatter.}

An antimatter system with a variable number of particles is characterised by
the equation
\begin{equation}
d\bar{U}_{a}=\bar{T}_{a}d\bar{S}_{a}+\bar{\mu}_{a}d\bar{N}_{a}%
\end{equation}
which remains the same as conventional as long as it is presented from the
perspective of the antiobserver (the other conjugate thermodynamic variables,
such as volume and pressure, can be added to this equation if needed). From
our perspective, this equation changes according to (\ref{TSU}). The
reversible equivalence requires preservation of mass, which demands that the
apparent and intrinsic numbers of particles composing antimatter are the same.
Hence, the apparent and intrinsic values of the number of particles (or moles)
$N$ and of the chemical potentials $\mu$ coincide:
\begin{equation}
N_{a}=\bar{N}_{a},\;\;\;\mu_{a}=\bar{\mu}_{a}%
\end{equation}
Similar relations can be drawn for other thermodynamic quantities such as
volume and pressure. Due to the observational symmetry, $\mu_{a}=\bar{\mu}%
_{a}(\bar{T}_{a},\bar{N}_{a},...)$ is the same function as $\mu_{m}=\mu
_{m}(T_{m},N_{m},...).$

We note first that at infinite temperatures $T_{m}=T_{a}=\bar{T}_{a}=\infty$,
the same quantities of matter and antimatter $N_{a}=N_{a}=N,$ which are kept
under the same conditions, are in both thermal and chemical equilibrium since
$\mu_{a}=\bar{\mu}_{a}(\infty,N,...)=\mu_{m}(\infty,N,...).$ If, however,
$T_{m}<\infty,$ then thermal equilibrium $T_{m}=T_{a}=-\bar{T}_{a}$ does not
imply that matter and antimatter under similar conditions are in chemical
equilibrium since $\mu_{a}=\bar{\mu}_{a}(-T,...)$ is generally different from
$\mu_{m}=\mu_{m}(T,...)$ for the same $T$. Hence, even if it might be possible
to put an antisystem in thermal equilibrium with a system by using negative
temperatures, this equilibrium does not extend to chemical stability between
matter and antimatter.

If a system and an antisystem are kept at the same intrinsic temperatures
$T_{m}=\bar{T}_{a}=T<\infty$ (and the same other conditions), then $\mu
_{a}=\bar{\mu}_{a}(T,N,...)=\mu_{m}(T,N,...)$. In this case, however, equality
of chemical potentials does not ensure chemical equilibrium since there is no
thermal equilibrium $T_{m}\neq T_{a}=-\bar{T}_{a}$.

Consider the following example: matter and antimatter are confined to the
system and antisystem correspondingly and can not mix (mixtures are considered
in the following subsection). Assume that matter and antimatter can react with
each other and can annihilate to produce radiation as reflected by the
following reactions\footnote{We use hypothetical reactions with neutrons $n$
and antineutrons $\bar{n}$ to illustrate our point. The neutron/antineutron
oscillation reaction $n\rightleftarrows\bar{n}$ \ (or similar reactions
converting antimatter into matter) is predicted by the 1st Sakharov condition
but has not been observed so far under current conditions prevailing in the
Universe.}
\begin{equation}
\text{(a)\ }n\rightleftarrows\bar{n},\;\;\;\text{(b) }n+\bar{n}%
\rightleftarrows2\gamma\label{reac}%
\end{equation}
These two reactions can be combined to result in conversion $2\bar
{n}\rightleftarrows2\gamma\rightleftarrows2n$; that is antimatter can be moved
from antisystem to system by radiation and converted to matter possessing the
same energy. Assuming that pressures are kept the same and constant in the
system and antisystem, this transition does not affect intrinsic temperatures
$T_{m}=\bar{T}_{a}$ and intrinsic entropy per particle (or per mole)
$s_{m}=\bar{s}_{a}$, where $s=S/N$. The total apparent entropy $S_{tot}%
=N_{m}s_{m}-N_{a}\bar{s}_{a}$ clearly increases as $N_{a}$ decreases and
$N_{m}$ increases. We see that CPT-invariant thermodynamics strongly favours
conversion of antimatter into matter (from observer's perspective), even if
the intrinsic properties of matter and antimatter are exactly the same. For an
antiobserver, whose time is going backward, this process seems as conversion
of matter into antimatter. \ Hence, from antiobserver's perspective,
thermodynamics favours conversion of matter into antimatter.

\subsection{Mixed matter and antimatter}

When matter and antimatter are mixed, they do not form semi-autonomous
thermodynamic systems with effectively independent directions of time. We
shall distinguish two cases: when matter dominates antimatter in the mixture
and when a 50:50 mixture is formed. Due to annihilations, radiation is also
present in the mixture but does not affect the direction of the thermodynamic time.

\subsubsection{Mixtures dominated by matter.}

Consider a mixture of many particles $N_{m}$ with very few antiparticles
$N_{a}$. The external degrees of freedom are entangled to produce a dominant
direction of thermodynamic time for the whole mixture. Since particles
dominate antiparticles in this mixture, this direction must be forward in
time. Presence of antiparticles may slightly increase the fluctuating
components of the time primer without any noticeable effect on the direction
of thermodynamic time. If particles and antiparticles are considered as
relatively simple mechanical or quantum systems, their thermodynamic behaviour
is primed by the mixture resulting in analogous thermodynamic properties
of\ the particles and antiparticles (the effect of thermodynamic interference
with the invariant properties of quantum systems is discussed in
Ref.\ \cite{K-PhysA}). The antiparticles, however, affect the thermodynamic
state of the mixture by causing annihilations.

A different situation appears when each particle and each antiparticle
represent an autonomous thermodynamic subsystem (i.e. having a large number of
comparable internal degrees of freedom, substantial ergodic mixing and
Kolmogorov-Sinai entropy \cite{Ergo-theory,K-PS2012T} dramatically amplifying
the effect of time priming). In this case, the thermodynamic time runs in
opposite directions within particles and antiparticles (and in the normal
direction for the whole mixture). While the thermodynamic subsystems can, in
principle, be placed in thermal equilibrium when intrinsic temperature of the
antiparticles is negative (this is possible when internal energy levels of
antiparticles are bounded --- see \cite{K-OTJ2012}) but this equilibrium is
typically unstable as discussed previously. Hence, antiparticle subsystems
should fall into their ground state (which is nevertheless subject to
fluctuations, significant for microscopic subsystems), while particle
subsystems remain in conventional thermal equilibrium with the mixture.
Although techniques based on evaluating statistical sums can not be used for
systems with negative heat capacity (see \cite{K-OTJ2012}), there is no
contradiction in using these techniques when antiparticle systems are in their
ground state. Let us see how the partition function $Z$ should be evaluated in
this case
\begin{gather}
Z=(Z_{0})^{N_{m}+N_{a}}\frac{\left(  Z_{m}\right)  ^{N_{m}}}{N_{m}!}%
\frac{\left(  Z_{a}\right)  ^{N_{a}}}{N_{a}!},\label{Z1}\\
Z_{m}=\sum_{i}\exp\left(  -\frac{E_{i}}{k_{B}T}\right)  ,\;\;Z_{a}=\exp\left(
-\frac{E_{0}}{k_{B}T}\right)  , \label{Z2}%
\end{gather}
where $Z_{0}$ is the partition function without considering internal degrees
of freedom, $Z_{m}$ is partition function for particle internal energy levels
and $Z_{a}$ is the partition function for antiparticles in their ground state.
Classical statistics is assumed in this example. Note that $Z_{a}$ can be made
unity ($Z_{a}=1$) without loss of generality by selecting $E_{0}$ as the
reference energy level (i.e. $E_{0}=0$). The difference in chemical potentials
of particle and antiparticles can be evaluated by using standard techniques
\cite{LL5}
\begin{align}
\Delta\mu &  =\mu_{a}-\mu_{m}=\left(  \frac{\partial A}{\partial N_{a}}%
-\frac{\partial A}{\partial N_{m}}\right)  _{T}\nonumber\\
&  =k_{B}T\left(  \ln\left(  \frac{N_{a}}{N_{m}}\right)  +f_{m}(T)\right)
\label{dmu}%
\end{align}
where $A=-k_{B}T\ln(Z)$ is the Helmholtz potential and $f_{m}(T)=\ln
(Z_{m}(T))\geq0$. Since $\Delta\mu>0$ thermodynamics favours conversion of
antimatter into matter under these conditions. The asymmetry of the chemical
potentials $\Delta\mu$ tends to be higher at higher temperatures. Note that
particles and antiparticles may have not only different chemical potentials
but also slightly different masses as they have different average energies.

The system containing particles $n$, antiparticles $\bar{n}$ and radiation
$\gamma$ which are engaged in reactions specified by (\ref{reac}) is now
considered. Reactions (\ref{reac}) require the following equilibrium
conditions (a) $\mu_{m}=\mu_{a}$ and (b) $\mu_{m}+\mu_{a}=2\mu_{\gamma}$.
Considering that photons should have zero chemical potential $\mu_{\gamma}=0$
\cite{LL5}, we conclude that $\mu_{m}=\mu_{a}=0$. Since, under conditions
specified above, $f_{m}$ is expected to be positive, equation (\ref{dmu})
indicates that particles dominate antiparticles $N_{m}>N_{a}$ --- this is
consistent with the major assumption of this subsection.

\subsubsection{The 50:50 mixture.}

The case of having 50\% particles and 50\% corresponding antiparticles in the
mixture is the most difficult case to analyse. The thermodynamic time can not
run in any direction due to matter/antimatter symmetry. This generally means
that only reversible processes can occur in this mixture and there is no
relaxation towards equilibrium state in any direction of time. Any process
that is irreversible forward or backward in time is impossible in this
mixture. The mixture can also contain a large or small fraction of radiation,
which does not affect the direction of time. CPT-invariant thermodynamics
provides a thermodynamic perspective for Sakharov's \cite{Sakharov1967}
hypothesis of CPT invariance of the Universe (see also \cite{K-M-2012}) and
changes interpretation of the third Sakharov condition listed in Section 2.
The 50:50 mixture of matter and antimatter, which presumably existed after the
initial singularity of the Big Bang, does not allow for gravitational
collapses (such as formation of black hole or disappearance of a white hole)
since these processes involve significant changes in entropy. This places
constraints on the properties of the Universe at its origin. 

Although there is no direction of thermodynamic time on average in this
mixture, there are fluctuations in the system. The thermodynamic time might
also fluctuate, moving slightly forward or backward due to a minor local
prevalence of one of the mixture components over the other. At this point
things get more complicated. The areas with a small excess of matter can be
seen as forming systems, while the areas with a small excess of antimatter
form antisystems. If all temperatures are infinite $T_{m}=T_{a}=\bar{T}%
_{a}=\infty$, then, according to the analysis of the previous subsection,
systems and antisystems can be considered to be in equilibrium with respect to
matter/antimatter conversion. If, however, the intrinsic temperatures are high
but finite, then (as also discussed in the previous subsection) thermodynamics
favours transfer of energy, volume and matter/antimatter from antisystems to
systems. This process can be seen as thermodynamic instability resulting in
systems (with the forward-directed thermodynamic time) taking over and
antisystems (with the backward-directed thermodynamic time) disappearing. In
antisystem regions, matter and antimatter can unmix forward in time reducing
further the volume occupied by antisystems. Matter needs to retain its leading
role over antimatter within the system regions, if the forward-time evolution
is to continue in these regions. The asymmetry of chemical potentials
(\ref{dmu}) stimulates conversion of antimatter into matter within the
systems. Note that the same mechanism converts matter into antimatter within
the antisystems but this happens in the forward direction of antiobserver's
time (i.e. backward in our time). Hence, from the observer's perspective,
antimatter is still converted into matter in this process.

\section{Testing invariant properties of thermodynamics}

In our world, which is exclusively populated by radiation and matter, the
CP-invariant and CPT-invariant versions of thermodynamics produce the same
predictions, are not experimentally distinguishable and can be treated as
different interpretations of the same physical theory. Specifically,
CP-invariant thermodynamics treats the temporal irreversibility as CPT
violation, while CPT-invariant thermodynamics sees CPT invariance as the
fundamental property of nature and relates the arrow of time to asymmetric
presence of matter and absence of antimatter in the universe. The CP- and CPT-
invariant versions of thermodynamics, however, represent different and
experimentally distinguishable physical theories in interpreting thermodynamic
properties of antimatter, which (at least in principle) is possible to produce
in high-energy colliders in statistically significant quantities
\cite{Nature2007,H2anti2010}.

On the face of the problem making an experimentally justified choice between
CP-invariant and CPT-invariant versions of thermodynamics is not too difficult
--- we just need to produce stochastically significant quantities of
antimatter (whose intrinsic thermodynamic time runs backward in our time
according to CPT-invariant thermodynamics) and insulate it from
time-generating interference of the surrounding matter. The antimatter system
must be sufficiently complex and interacting to possess significant
Kolmogorov-Sinai entropy and amplify its intrinsic (spontaneous) time primer.
If the physical reality is compliant with the CPT invariance, antimatter
assembled in quantities exceeding certain critical level would change the
direction of its thermodynamic time. Note that under these conditions,
explosion of antimatter into a burst of coherent light directed at surrounding
matter becomes thermodynamically possible since antimatter has negative
apparent temperatures in our world. Besides creating and containing antimatter
in sizeable quantities, the major difficulty with this experiment is that we
do not know the mechanism of time priming, its amplitude and physical
properties. We thus can not be certain about relative contributions of induced
and intrinsic time priming in a particular experiment. Since the exact
mechanism of time priming remains unknown, we cannot examine its invariant
properties directly, without assembling thermodynamically significant
quantities of antimatter.

Recent experiments in high-energy accelerators, indicate that thermodynamics
might be relevant to very small scales (within a hadron) and high energies
\cite{MuB2011}. Baryons (protons and neutrons) seem to contain myriads of
appearing and annihilating gluons, quarks and antiquarks, as specified by
so-called parton distribution \cite{Martin2001}. Collisions involving protons
and nuclei produce various particles and antiparticles with distributions
strongly resembling a thermodynamic equilibrium. This equilibrium is
characterised by a number of parameters, including $\mu_{B}$ --- the chemical
potential associated with the baryon number $B$, so that baryons (i.e. protons
and neutrons having $B=+1$) and antibaryons (antiprotons and antineutrons
having $B=-1$) have different chemical potentials \cite{MuB2011}. This seems
to confirm the CPT-invariant version of thermodynamics (since in CP-invariant
thermodynamics, properties of matter and antimatter must be the same and
$\mu_{B}=0$) but, in our opinion, this is probably not the case. In the
absence of reactions violating $B$, true chemical potential associated with
the baryon number is not revealed\footnote{This can be compared this with
chemical reactions, where elements can be assigned arbitrary chemical
potentials --- in absence of nuclear reactions, true chemical potentials of
the elements remain unknown.}. Hence, $\mu_{B}$ is likely to be an effective
quantity reflecting initial conditions, although it does seem that
thermodynamic equilibrium is achieved in these experiments with respect to the
other parameters.

In spite of the difficulties in creating significant quantities of antimatter,
thermodynamic antisystems, may already exist in high-energy experiments. Here
we again refer to the parton distribution \cite{Martin2001}, describing
behavior of a large number of quarks, antiquarks and gluons in protons,
neutrons and other hadrons. The parton distribution is biased towards two or
three valent quarks that determine properties of the particles. It is likely
that at low temperatures (that can still be very high according to our common
understanding of what is hot and what is cold) baryons are in their quantum
ground state. In this case thermodynamics does not play a role in internal
properties of these systems. If the temperature becomes extremely high (of
order of $10^{12}K$) quarks, antiquarks and gluons break the confinement of
hadrons and form the so-called quark-gluon plasma \cite{MuB2011}, which is a
thermodynamic object and thus must be in a quantum mixed state. It is likely
that at intermediate temperatures, such complex particles still remain intact
but are in a mixed state and display a thermodynamic behaviour. Hence, we
might be able to treat baryons as thermodynamic systems (here we refer to the
particles themselves and not to various large thermodynamic systems that can
be formed by these particles), implying existence of small thermodynamic
antisystems in our world in form of antibaryons.

Let us investigate thermodynamic implications of this assumption. We note
first that the spontaneous time primer is then likely to exist within baryons
and antibaryons: superposition of quantum states in neutrons would tend to
lose coherence (de-coherence) while mixed states in antineutrons would tend to
become pure superpositions (i.e. en-coherence in our time). The analysis of
previous section indicates that in the environment dominated by matter,
antisystems tend to remain in their ground states (this does not exclude some
level of thermodynamic fluctuations) while systems reach a quantum mixed state
with the temperature similar to that of the environment. Since a large
fraction of energy of protons and neutrons is in motions of quarks and gluons,
a mass defect between particle and antiparticle is possible at a sufficiently
high temperature. The particle possessing higher internal thermal energy
should have a somewhat higher mass as compared to their antiparticle
counterparts. Let us turn to experimental evidence. While the equivalence of
the masses of most particles and antiparticles (e.g. electrons and positrons,
protons and antiprotons) has been confirmed with a very small relative error
of around 10$^{-6}$-10$^{-7}$ per cent \cite{PDG2012}, experimental
confirmation of equivalence of the neutron and antineutron masses is subject
to a greater uncertainty. Measurements indicate that the neutrons are heavier
than antineutrons by a small (and rather uncertain) fraction, which is
estimated in Ref.\cite{PDG2012} as 0.009$\pm$0.006\%. CPT-invariant
thermodynamics does not see this kind of discrepancies as violations of the
fundamental CPT symmetry but tends to relate them to thermodynamic
interference with the measurements resulting in different intrinsic
temperatures of neutrons and antineutrons. Neutrons have higher internal
thermal energy than antineutrons in our matter-populated environment, as
expected from the analysis of the previous section. If assumptions about
thermodynamic state of neutrons introduced in this section are correct, this
difference should decrease, when neutrons are contained and cooled down before
the experiments. Unlike the CPT-invariant version of thermodynamics, the
CP-invariant version specifies exactly the same thermodynamic properties for
both neutrons and antineutrons.

\section{Discussion and conclusions}

Assuming a fundamental similarity of the intrinsic properties of matter and
antimatter, conventional thermodynamics can be extended to include antimatter
in two different ways: CP-invariant and CPT-invariant. Due to the
time-directional nature of thermodynamics, its CP-invariant and CPT-invariant
versions can not be valid at the same time. Philosophically, CPT-invariant
thermodynamics connects two major asymmetries in nature --- the observed
direction of time, and the relative abundance of matter and\ absence of
antimatter, while CP-invariant thermodynamics sees these issues as separate.
In the absence of appreciable quantities of antimatter in our world, it is
very difficult to determine which version of thermodynamics is not only
logically possible but also real: both versions give the same predictions for
the world dominated by matter.

The difference between CP- and CPT-invariant thermodynamics, however, is not
limited to philosophical interpretations. CPT-invariant thermodynamics
indicates that nature has difficulties in assembling substantial quantities of
antimatter due to its thermodynamic antagonism with matter. There is no
thermodynamic antagonism between matter and antimatter according to
CP-invariant thermodynamics. Collecting statistically significant quantities
of antimatter at a sufficient density and insulating it from the dominant time
priming influence of the environment should result, according to CPT-invariant
thermodynamics, in changing the thermodynamic arrow of time and a
thermodynamically unstable state associated with negative apparent
temperatures. In this state, transfer of energy and mass from antimatter to
matter (forward in our time) is favoured by thermodynamics. Yet these
differences in apparent behaviour are consistent with the fundamental
similarity between intrinsic properties of matter and antimatter. CP-invariant
thermodynamics, on the contrary, does not expect any thermodynamic
instabilities in such experiments and suggests that the apparent thermodynamic
properties of matter and antimatter are the same.

We should assume that spontaneous time-priming processes are of very small
magnitude and, thus, are not directly detectable in conventional experiments.
This is plausible since, in spite of the clear presence of the arrow of time
(presumably everywhere, including in remote and isolated systems), we still do
not know its exact mechanism. The presence of some scattered elementary
antiparticles in a world dominated by matter does not change the direction of
thermodynamic time. If, however, baryons (neutrons and protons) can be treated
as thermodynamic systems due to the complexity of their parton distributions,
then CP-invariant thermodynamics requires exact similarity of the chemical
potentials of baryons and antibaryons, while CPT-invariant thermodynamics
predicts differences in effective chemical potentials between baryons and
antibaryons in high-temperature environments.

While distinguishing two extensions of conventional thermodynamics
(CP-invariant and CPT-invariant) can be quite difficult in the real world,
which we are familiar with and which is essentially free from antimatter, this
may change as more and more substantial quantities of antimatter are produced
in experiments by high-energy accelerators \cite{Nature2007,H2anti2010}.

\acknowledgements{Acknowledgements} The authors acknowledge funding by the
Australian Research Council.

\bibliographystyle{mdpi}
\bibliography{Law3}

\appendix

\section*{APPENDIX: Boltzmann's time hypothesis, the direction of time and the
time primer}

While this work is mostly interested in macroscopic aspects of extending
thermodynamics from matter to antimatter, it is difficult to stay within
macroscopic treatment of conventional thermodynamics when particles and
antiparticles form mixtures rather than (autonomous systems and antisystems).
These cases are most complicated and need a working concept for the direction
of thermodynamic time. This and other associated issues are briefly discussed
below in context of the Boltzmann time hypothesis linking our perceived
direction of time to the second law of thermodynamics. The temporal boundary
conditions imposed on the Universe affect the direction of time directly and
through the properties of matter.

\begin{figure}[h]
\begin{center}
\includegraphics[width=10cm,page=2,trim=2cm 3cm 6cm 4cm, clip ]{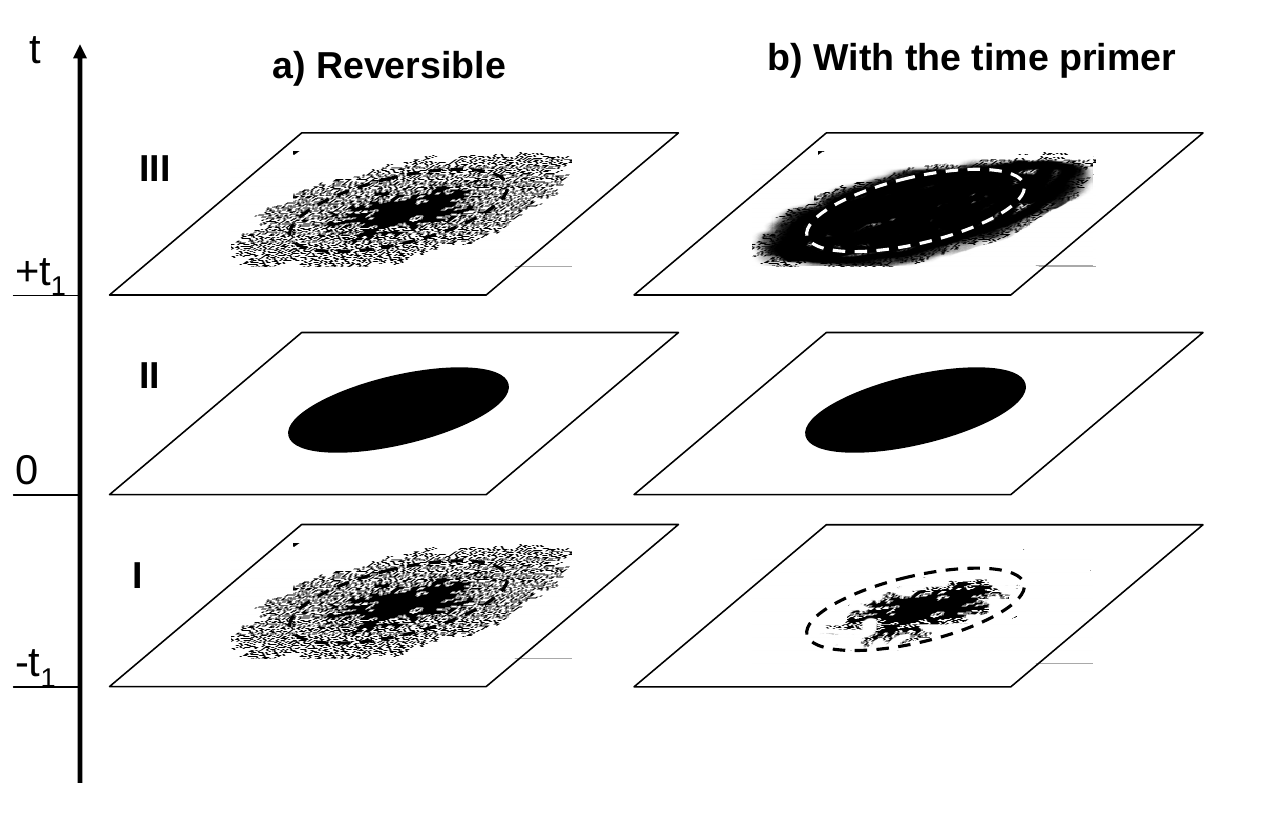}
\caption{Schematic for evolution of a selected volume in the phase space in case of a) reversible dynamics and b) under influence of the timer primer. }
\label{fig2}
\end{center}
\end{figure}

From the perspective of the ergodic theory, increase of entropy in dynamic
systems of large dimensions is associated with ergodic mixing
\cite{Ergo-theory,K-PS2012T}. This perspective is based on classical
mechanics, which is time-reversible and preserves volume measures in the
extended phase space (whose dimension is very large and determined by the
overall number of microscopic degrees of freedom in the system). Note that
quantum mechanics possesses similar properties linked to the unitarity of
quantum evolutions. Ergodic mixing can be understood through analogy between
evolution of volumes in the extended phase space with mixing of fluids. This
analogy was first introduced by Gibbs in his fundamental work \cite{Gibbs1902}%
, which laid the foundations of statistical physics. Ergodic mixing can be
illustrated by Figure \ref{fig2}, which depicts the extended phase space of a
dynamic system. The ensemble of states is initially (at $t=0$) confined to the
black region in Figure \ref{fig2}(IIa). The evolution of the system changes
the shape of the region but does not alter its volume measure so that after a
sufficiently long period of time $t_{1}$, the trajectories become densely
distributed in a larger and larger segment of the domain. Refer to Figure
\ref{fig2}: the black areas have exactly the same measure (i.e. the same
number of black pixels) in IIa and IIIa but the segment effectively occupied
by the black spot is larger in IIIa. The increasing volume measure of this
segment is conventionally associated with an increasing entropy, as required
by the second law (in dynamic systems, the increase of the coarsened phase
volume in time is characterised by the \textit{Kolmogorov-Sinai dynamic
entropy }\cite{KSE1999}). This logic, however, encounters severe problems when
states of the system are considered back in time \cite{PenroseBook}. Since the
evolution of a dynamic systems is time-reversible, it can be extended back to
negative times and a state of the system similar to the state shown in IIIa
should also be expected at $t=-t_{1}$ \ (see Figure \ref{fig2}, Ia). How can
time-symmetric evolution (compare Ia and IIIa in Figure \ref{fig2}) be
consistent with the second law?

The resolution of this apparent paradox is commonly related to the set of
ideas, which we call here \textit{priming of time. }We use this term to refer
to a generic process that triggers or primes the mechanism of enacting the
direction of thermodynamic time.\textit{ }This process causes a diffusion-like
increase of the black volume into the white region so that the state of the
system at $t=t_{1}$ looks as shown in Figure \ref{fig2}(IIIb). The time primer
does not refer to a specific theory or hypothesis but is seen as a violation
of classical mechanics and/or quantum mechanics; although the magnitude of
these violations is so small that it is very difficult to detect the time
primer directly. The influence of the time primer on the Universe is
nevertheless profound. Due to ergodic mixing, the area of the black/white
interface in Figure \ref{fig2}(IIIa) becomes so large that even very small
coarsening of the distribution results in a large increase of the black
volume. Hence entropy increases forward in time. The time primer is not
time-symmetric and when we move back in time to $t=-t_{1}$, it acts to
decrease the black volume as depicted in Figure \ref{fig2}(Ib). Hence entropy
reduces back in time. According to Boltzmann's time hypothesis, our perceived
direction of time is that of the thermodynamic time, which is linked to the
second law of thermodynamics. The direction of thermodynamic time is
determined by the temporal direction of the time primer.

In spite of the overwhelming evidence for directional asymmetry of time around
us, the exact physical nature of the time primer still remains a mystery. In
quantum mechanics, the time primer is linked with the process of decoherence,
which is expected to violate unitarity of quantum evolutions.\ Although time
symmetry is broken in quantum mechanics by weak interactions, there is no
evidence that the known T-violating processes are the ones enacting the
direction of time. A few theories (see, for example,
\cite{Beretta2005,QTreview}) are related to the process of quantum
decoherence, which increases the number of accessible quantum states. Penrose
\cite{PenroseBook} believes that decoherence is linked to interactions of
quantum mechanics and gravity (although gravity, classical or relativistic, is
conventionally seen as time-symmetric). We, however, do not link the time
primer to any specific theory or mechanism. In principle, the time primer may
be a) a spontaneously generated property of matter and antimatter or b) can be
induced by interactions with environment or, possibly, with a special
time-generating field. It is likely that generation of the time primer
combines the \textit{spontaneous }(a) and \textit{induced} (b) mechanisms. In
line with adopted terminology, spontaneous time priming can also be called
\textit{intrinsic}. This means that the temporal irreversibility can be
initially generated within matter (and also antimatter) and then propagate
through interactions (for example, interactions by radiation). Even very small
interactions, which have magnitudes below the detection limits, can be
sufficient to tip entropy increase towards the common direction of time.

\begin{figure}[h]
\begin{center}
\includegraphics[width=8.5cm,page=3,trim=6cm 3cm 9cm 5cm, clip ]{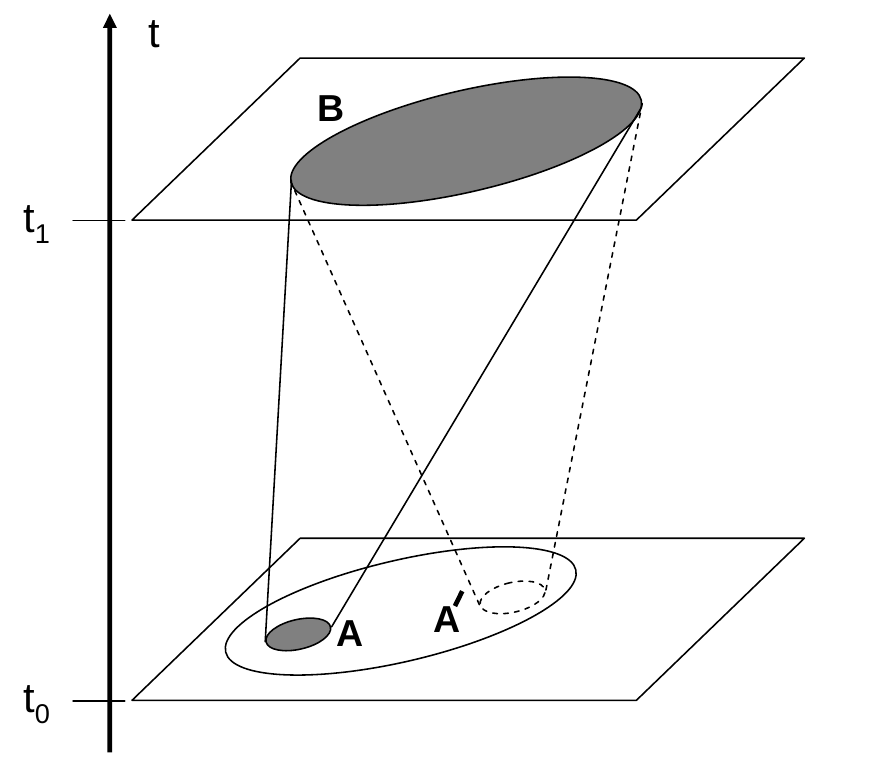}
\caption{ Increase of the phase space volume in irrevesible evolutions and causality.}
\label{fig3}
\end{center}
\end{figure}

The asymmetric nature of the time primer enacts the second law, which in its
turn enacts causality: we see that the future is determined by the past but
not vice versa. Indeed, consider evolution of the phase volume schematically
shown in Figure \ref{fig3}. If the initial state of the system is known as,
say, area A at $t=t_{0}$, the state of the system at any following moment can
be predicted: for example, it is area B at $t=t_{1}.$\ The converse statement
is, however, incorrect due to action of the second law: knowing that the
system is in state B at $t=t_{1}$ does not allow us to conclude that it was in
state A at $t=t_{0}$, since the system state at $t=t_{0}$ could well be
A$^{\prime},$ or many other states determined by various possible contractions
of the phase volume measure. While moving backward in time, the choice between
A and A$^{\prime}$ is seen as a random event. We interpret this by saying that
A causes B but B is not a cause for A. The causality principle allows to treat
mechanical evolutions (classical or quantum) as time-reversible, while
replacing the temporal irreversibility by requirement of setting the initial
conditions before (and not after) the process under consideration. This
principle provides a great simplification and works really well in many
respects. The philosophical aspects of time arrow and causality are discussed
in a book by Price \cite{PriceBook} and in a number of more recent
publications \cite{time2003,time2010b,time2010a}.

The causality principle has another important implication: a non-equilibrium
state must have its cause in the past but not in the future. For example,
footsteps on a beach will eventually reach equilibrium with the sand and
disappear (no special effort is needed for this) but the footsteps could not
appear without a reason (i.e. someone walking across the beach) as this would
contradict to the second law. A system that evolves very slowly towards its
equilibrium (a photograph, for example) is a testimony of a distant past when
its initial state was created by an external disturbance. The main implication
of causality is that we can remember the past and can predict the future but
we cannot remember the future and cannot retrodict the past (pure retrodiction
is ill-posed since many different pasts correspond to nearly the same future
but partial retrodiction combined with partial memory records can be much more feasible).

The time primer does not have to be a fully deterministic process and the
phase volume may fluctuate slightly from point to point, provided the overall
trend of the phase volume to increase forward in time is dominant. We,
nevertheless, may observe some randomness forward in time. The forward-time
randomness can be seen as a minor fluctuation of the time primer process,
which results in a small local reduction of the phase volume. For example a
rolling coin will finish on one of its sides, but it is impossible to predict
if this is going to be ''heads'' or ''tails''. When the outcome becomes known,
the reduction of uncertainty in this single bit (i.e. heads or tails) does not
cause a global entropy decrease since the falling coin dissipates energy and
this is accompanied by a large increase in molecular entropy. Another example
of forward-time randomness is collapse of the quantum wave function during
measurements, which should be accompanied by offsetting entropy increase
somewhere in the measuring apparatus. 
\end{document}